\documentclass{article}
\usepackage{graphicx}
\usepackage{color}
\usepackage[separate-uncertainty=true]{siunitx}
\usepackage{amsfonts, amssymb} 
\usepackage{url}
\usepackage{caption}
\usepackage[minnames=1,maxnames=5,style=numeric-comp,sorting=none,backend=bibtex]{biblatex}
\usepackage[super]{nth}
\usepackage[version=3]{mhchem}
\usepackage{float}

\bibliography{sidemotion.bib}
 
\newcommand{\fp}{Fabry-P\'{e}rot }

\begin{document}
\title{Upper Limit to the Transverse to Longitudinal Motion Coupling of a Waveguide Mirror}
\author{S\,Leavey$^{1}$ \and B\,W\,Barr$^{1}$ \and A\,S\,Bell$^{1}$ \and N\,Gordon$^{1}$ \and C\, Gr\"{a}f$^{1}$ \and S\,Hild$^{1}$ \and S\,H\,Huttner$^{1}$ \and E-B\,Kley$^{2}$ \and S\,Kroker$^{2}$ \and J\,Macarthur$^{1}$ \and C\,Messenger$^{1}$ \and M\,Pitkin$^{1}$ \and B\,Sorazu$^{1}$ \and K\,Strain$^{1}$ \and A\, T\"{u}nnermann$^{3}$}
\date{}

\maketitle

Correspondence: \url{s.leavey.1@research.gla.ac.uk}

\begin{abstract}
Waveguide mirrors possess nano-structured surfaces which can potentially provide a significant reduction in thermal noise over conventional dielectric mirrors. To avoid introducing additional phase noise from motion of the mirror transverse to the reflected light, however, they must possess a mechanism to suppress the phase effects associated with the incident light translating across the nano-structured surface. It has been shown that with carefully chosen parameters this additional phase noise can be suppressed. We present an experimental measurement of the coupling of transverse to longitudinal displacements in such a waveguide mirror designed for \SI{1064}{\nano\meter} light. We place an upper limit on the level of measured transverse to longitudinal coupling of one part in seventeen thousand with 95\% confidence, representing a significant improvement over a previously measured grating mirror.
\end{abstract}
1. SUPA, School of Physics and Astronomy, The University of Glasgow, Glasgow, G12\,8QQ, UK\\
2. Friedrich-Schiller-University, Abbe Center of Photonics, Institute of Applied Physics, Max-Wien-Platz 1, 07743 Jena, Germany\\
3. Fraunhofer Institute of Applied Optics and Precision Engineering, Albert-Einstein-Str. 7, 07745 Jena, Germany

\section{Introduction}
\label{sec:intro}

Major upgrades to the worldwide network of gravitational wave detectors are
currently under way. New designs for the Advanced LIGO \cite{Harry2010},
Advanced
Virgo \cite{Avirgo2009}, KAGRA \cite{Somiya2012} and GEO-HF \cite{Willke2006}
detectors will provide unmatched ability to detect gravitational waves in the
audio spectrum.
At their most sensitive frequencies, these detectors are expected to be limited
by Brownian thermal noise arising from the reflective coatings on the
detectors' test masses \cite{Levin1998, Nakagawa2002, Harry2002, Crooks2002}. In order to help mitigate this limitation beyond the
next generation of detectors, efforts are under way to develop mirror coatings
with lower thermal noise \cite{Flaminio2010, Bassiri2013}.

In the case of Advanced LIGO, each end test mass (ETM) consists of a substrate with 19
pairs of sub-wavelength coatings which produce a transmission of
\SI{5}{ppm} for \SI{1064}{\nano\meter} light \cite{Dannenberg2009}. Each layer
within this stack contributes to the overall thermal noise \cite{Harry2002, Crooks2002}. The approach taken by Levin to calculate the thermal noise of mirrors \cite{Levin1998} shows that mechanical loss at the front surface of a mirror contributes more to the Brownian noise level than loss from an equivalent volume in the substrate. Additionally, typical coating materials tend to exhibit mechanical loss orders of magnitude higher than typical substrate materials \cite{Harry2002, Crooks2002}. For these reasons particular attention is being given to the reduction of coating thermal noise to improve the sensitivity of future detectors.

One strategy, to be applied for example in KAGRA, is to cool the mirrors to
cryogenic temperatures. While this can potentially reduce the thermal noise of the mirrors \cite{Uchiyama2012}, the application of cryogenic mirrors requires new infrastructure, different choices of mirror substrate and coating materials and poses the challenge of heat extraction from the mirror without spoiling its seismic isolation and thermal noise performance. Efforts in the application of cryogenics are also under
way to identify suitable substrate and coating materials for ET-LF, the low
frequency interferometer as part of the proposed Einstein Telescope
\cite{Punturo2010, Martin2010, Hild2011, Abernathy2011}.

\begin{figure}
  \begin{center}
    \includegraphics[width=0.7\columnwidth]{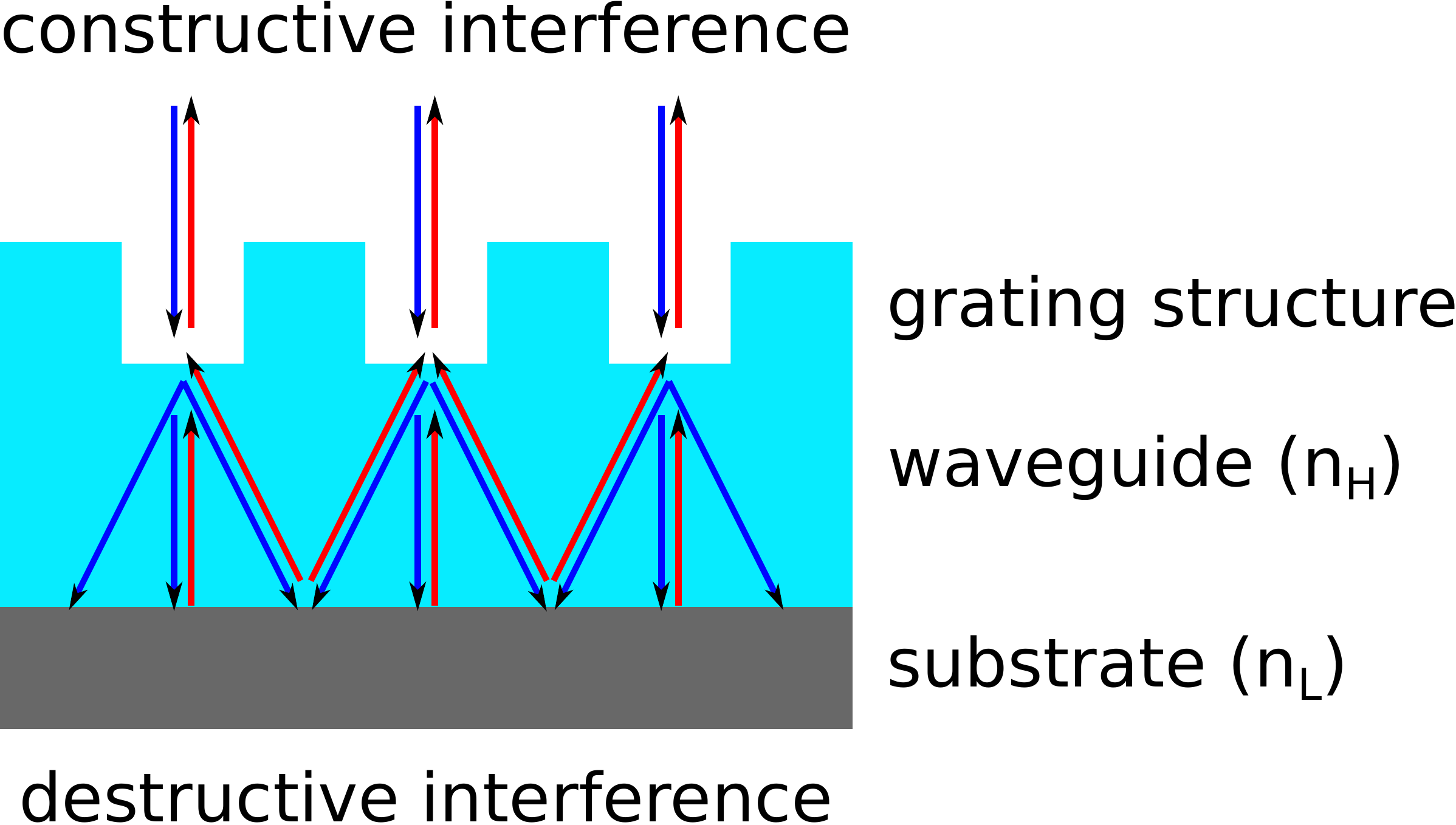}
    \caption{Propagation of light within a waveguide mirror. The grating
and waveguide layers have refractive index $n_H$, and sit atop a substrate of
refractive index $n_L$. Blue arrows represent incident light and red arrows
represent reflected light. In realisations of waveguide mirrors such as this, a
thin etch-stop layer is placed between the grating and waveguide layers to
assist fabrication \cite{Friedrich2011}.}
    \label{fig:waveguide_reflection}
  \end{center}
\end{figure}

Apart from using different coating materials \cite{Granata2013, Cole2013} or different beam shapes \cite{Mours2006, DAmbrosio2004, Bondarescu2006} such as with LG33 modes \cite{Sorazu2013}, another potential approach is to
utilise waveguide mirrors (WGMs) \cite{Brueckner2008, Brueckner2009, Brueckner2010, Friedrich2011}. These
mirrors can possess high reflectivity
at a wavelength determined by their structure. In contrast to conventional
dielectric mirrors, mirrors possessing waveguide coatings can exhibit high
reflectivity without requiring multiple stacks
\cite{Bunkowski2006}. A waveguide coating instead presents incident light
with a periodic grating structure of high refractive index material $n_H$ on top
of a substrate with low refractive index $n_L$ (see Figure
\ref{fig:waveguide_reflection}). Light is forced into a single reflective
diffraction order, the \nth{0}. In transmission, only the \nth{0} and
\nth{1} diffraction orders are allowed as long as the condition in Equation~\ref{eq:grating_equation} for the grating period, $p$; and the light's wavelength in vacuum, $\lambda$, is fulfilled \cite{Brueckner2008}. The light diffracted into the \nth{1} order undergoes total internal reflection at the substrate boundary where it excites resonant waveguide modes. Light leaving the waveguide then contains a \SI{180}{\degree} phase shift with respect to the \nth{0} order transmitted light, causing destructive interference such that most of the incident light is reflected \cite{Sharon1997}.

\begin{equation}
  \frac{\lambda}{n_{H}} < p < \frac{\lambda}{n_{L}}
  \label{eq:grating_equation}
\end{equation}

A recent set of calculations by Heinert \emph{et al.}~\cite{Heinert2013}
showed that a suitably optimised WGM can provide a reduction in coating thermal noise amplitude of a factor of 10 at cryogenic temperature compared to mirrors employed in Advanced LIGO.

\begin{figure}
  \begin{center}
    \includegraphics[width=0.5\columnwidth]{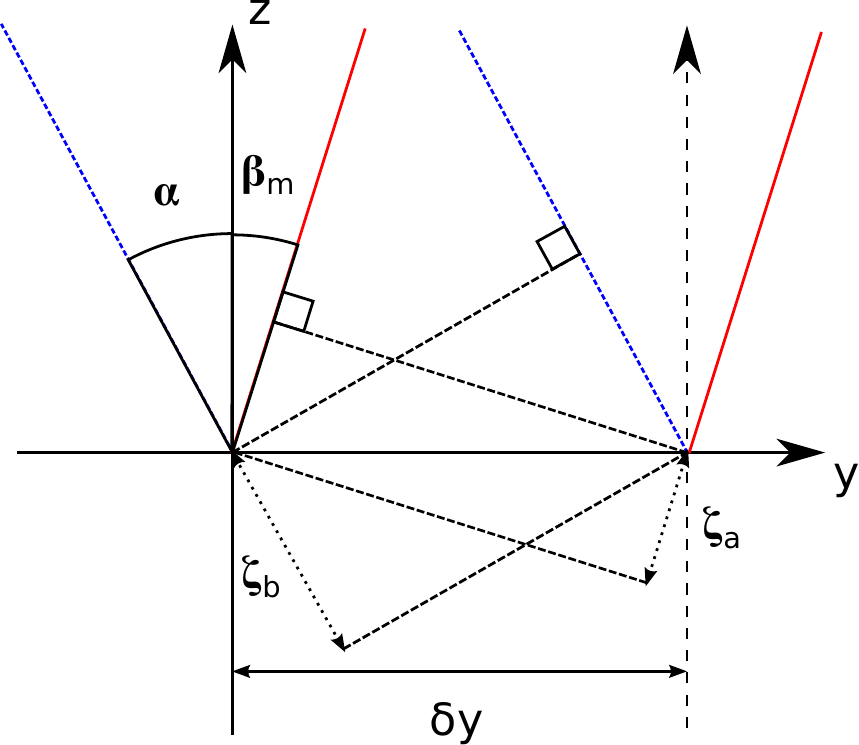}
    \caption{Optical path length changes $\zeta_a$ and $\zeta_b$ due to
transverse motion of a Littrow grating. Incident light diffracted into a
different order undergoes a path length change $\delta l_L =
\zeta_a + \zeta_b$.}
    \label{fig:grating_propagation}
  \end{center}
\end{figure}

Previous efforts to demonstrate grating structures as alternatives to dielectric
mirrors have identified phase noise in the light reflected from the grating not otherwise present in dielectric mirrors \cite{Wise2005, Freise2007}. This effect arises from transverse motion of grating mirrors with respect to the incident light. Incident light at angle $\alpha$ is reflected into the m\textsuperscript{th} diffraction order, exiting at angle $\beta_m$ (see Figure \ref{fig:grating_propagation}). The change in path length $\delta l_L$ between the reflected and incident light is then
\begin{equation}
  \delta l_L = \zeta_a + \zeta_b = \delta y
  \left( \sin{\alpha} + \sin{\beta_m} \right),
\end{equation}
where $\zeta_a$ and $\zeta_b$ represent the relative optical path length of each
depicted ray.
The phase modulation induced in the light reflected from the WGM is proportional to Fourier frequency with a \SI{90}{\degree} phase lead over the transverse motion \cite{Barr2011}. The noise added to the reflected light can be enough to mitigate the improvement in coating thermal noise, as witnessed in a study of \nth{2}
order Littrow gratings \cite{Barr2011}. Although WGMs also possess gratings, the
resonant waveguide structure has been shown in simulations by Brown \emph{et al.} to be invariant to transverse to longitudinal coupling \cite{Brown2013}.

\begin{table}
  \begin{center}
    \begin{tabular}{|l|l|}
      \hline
      \textbf{Parameter}    & \textbf{Value}          \\ \hline
      Materials             & \ce{SiO_2}, \ce{Ta_2O_5}, \\
                            & \ce{Al_2O_3} \\ \hline
      Design $\lambda$      & \SI{1064}{\nano\meter}   \\ \hline
      Grating depth         & \SI{390}{\nano\meter}   \\ \hline
      Waveguide depth       & \SI{80}{\nano\meter}    \\ \hline
      Etch stop depth       & \SI{20}{\nano\meter}    \\ \hline
      Grating period        & \SI{688}{\nano\meter}   \\ \hline
      Fill factor           & 0.38                    \\ \hline
      Reflectivity          & 96\% \\ \hline
    \end{tabular}
    \caption{Design parameters of the WGM produced by Friedrich-Schiller Jena
for the experiment to measure transverse to longitudinal coupling. It is similar to the one used in \cite{Friedrich2011}, with increased reflective surface area.}
    \label{tab:waveguide_parameters}
  \end{center}
\end{table}

\begin{figure}
  \begin{center}
    \includegraphics[width=0.5\columnwidth]{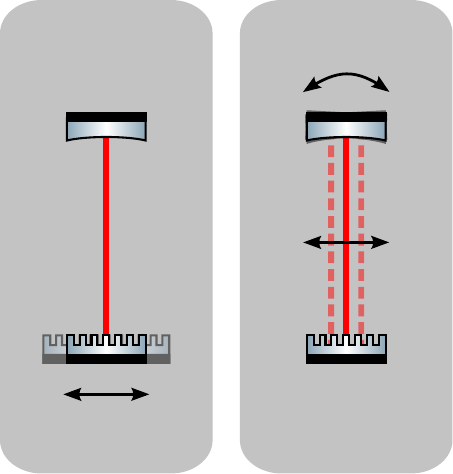}
    \caption{Two ways in which light can be scanned across the surface of the WGM. The left panel shows the effect of WGM motion with respect to a static beam, while the right panel shows the effect of light beam motion (due to rotation of the cavity mirror opposite the WGM) with respect to a static WGM. The latter effect is the one primarily considered in this article.}
    \label{fig:waveguide_scanning}
  \end{center}
\end{figure}

There are two mechanisms by which grating mirrors can couple transverse motion
into longitudinal phase changes (see Figure \ref{fig:waveguide_scanning}). The
first is through transverse motion of the grating, which can in principle be minimised with appropriate suspension design. The second mechanism is the coupling of changes in the opposite cavity mirror's alignment into the spot position on the grating
mirror. This effect is of particular importance to gravitational wave
observatories, where longer arm lengths can increase its detrimental impact.
For this reason the second mechanism is considered in more detail in this work.

In order to quantify its transverse coupling, a WGM was produced in collaboration with Friedrich-Schiller University Jena, Germany (see Table \ref{tab:waveguide_parameters} for its properties). It was designed for light of wavelength \SI{1064}{\nano\meter}, and consisted of an etched grating structure on top of a waveguide layer, both tantala, on a silica substrate. This article details an experiment carried out to measure its transverse coupling level.

\section{Experiment}

The fabricated WGM was used as the input coupler for a \fp cavity, held on
resonance using the Pound-Drever-Hall (PDH) technique \cite{Drever1983}. The error signal provided by the PDH technique represents changes in cavity length, and this can be fed back to the laser's frequency \emph{via} a frequency stabilisation servo.

\subsection{Cavity Length Signals}
\label{sec:lengthsignals}

A non-zero WGM transverse to longitudinal coupling, $\omega_1$, produces a phase shift on the reflected light. This manifests itself as an effective change in cavity length, $\delta l_W$, as the laser light is scanned across its grooves by a rotation of the ETM:
\begin{equation}
  \delta l_W \left( \theta, \kappa, \omega_1 \right) = \theta \kappa \omega_1,
  \label{eq:wgm_length_change}
\end{equation}
where $\theta$ is the ETM's rotation angle and $\kappa$ is the cavity's coefficient of ETM rotation to transverse WGM spot motion.

\begin{figure}
  \begin{center}
 
\includegraphics[width=0.5\columnwidth]{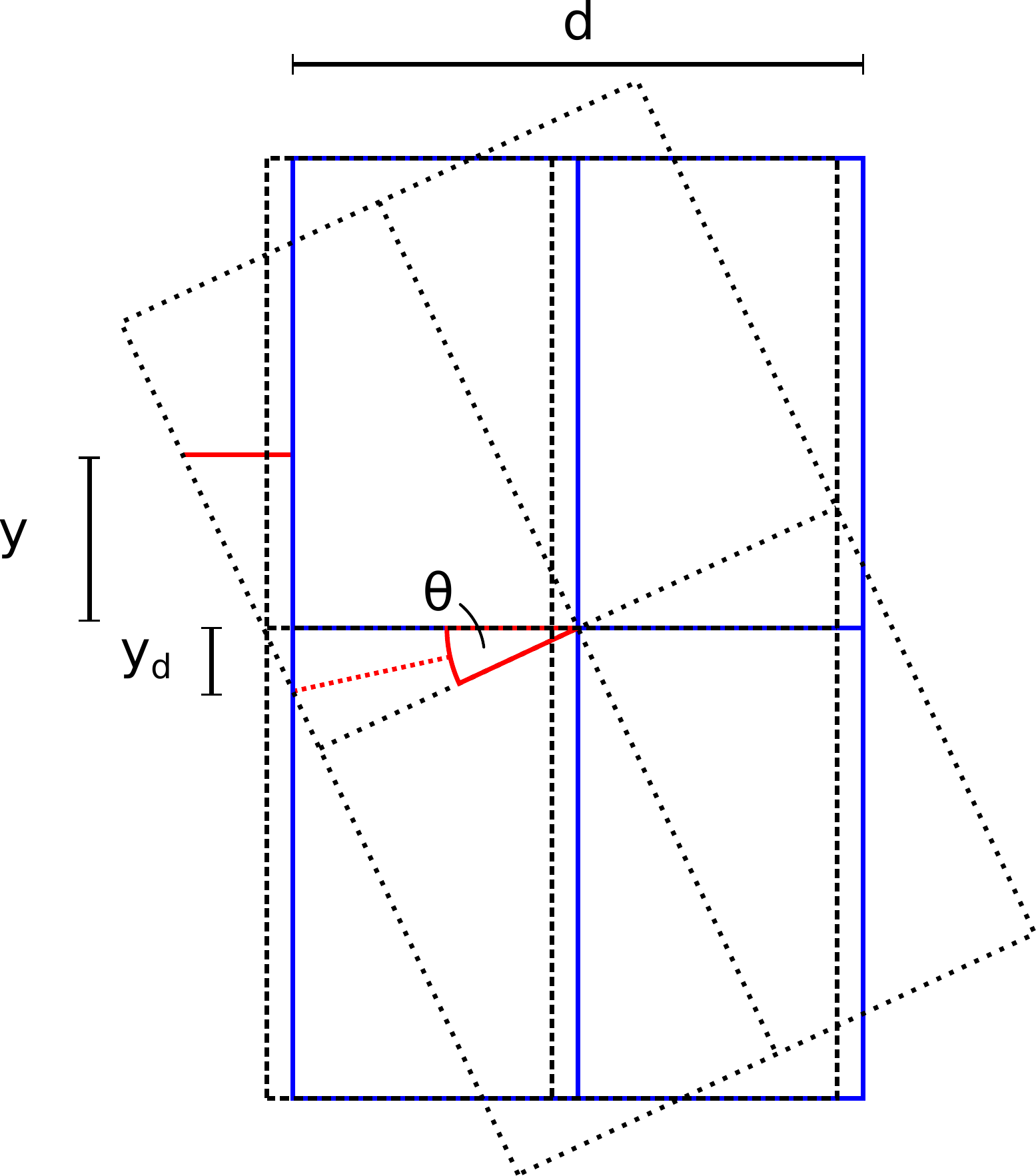}
    \caption{Geometrical ETM longitudinal effects. For a given rotation $\theta$ and spot
centre position offset $y$, the (longitudinal) position change in the surface of
the mirror (show in blue) as seen by the reflected light is approximately
$y \theta + \frac{d}{4} \theta^2$. The straight, solid red line in the figure
shows this longitudinal change.}
    \label{fig:mirror_longitudinal_effect}
  \end{center}
\end{figure}

Additional cavity length changes are also produced \emph{via} two geometrical effects (see Figure \ref{fig:mirror_longitudinal_effect}). The first effect, $\delta l_s$, is due to the position of the beam with respect to the centre of the mirror's surface. For a rotation $\theta$, a beam offset from the centre of the mirror by a displacement $y$ will receive a change in (longitudinal) path length of
\begin{equation}
  \delta l_s \left( y, \theta \right) = y \tan{\theta} \approx y \theta
  \label{eq:offset_effect}
\end{equation}
for small angles. The second effect, $\delta l_d$, is due to the depth $d$ of the mirror,
proportional to the rotation angle $\theta$. The position of the centre of the mirror with respect to the zero rotation case, $y_d$, is then
\begin{equation}
  y_d \left( d, \theta \right) = \frac{d}{2} \tan{\frac{\theta}{2}} \approx \frac{d}{4} \theta,
\end{equation}
and the change in path length this causes is
\begin{equation}
  \delta l_d \left( d, \theta \right) = y_d \tan{\theta} \approx \frac{d}{4} \theta^2.
\end{equation}
The total longitudinal effect $\delta l_E$ caused by the
rotation of the ETM is therefore
\begin{equation}
\delta l_E \left(y, \theta, d \right) = \delta l_s + \delta l_d \approx y \theta + \frac{d}{4} \theta^2.
\label{eq:etm_length_change}
\end{equation}

\begin{figure}
  \begin{center}
    
\includegraphics[width=1.2\columnwidth]{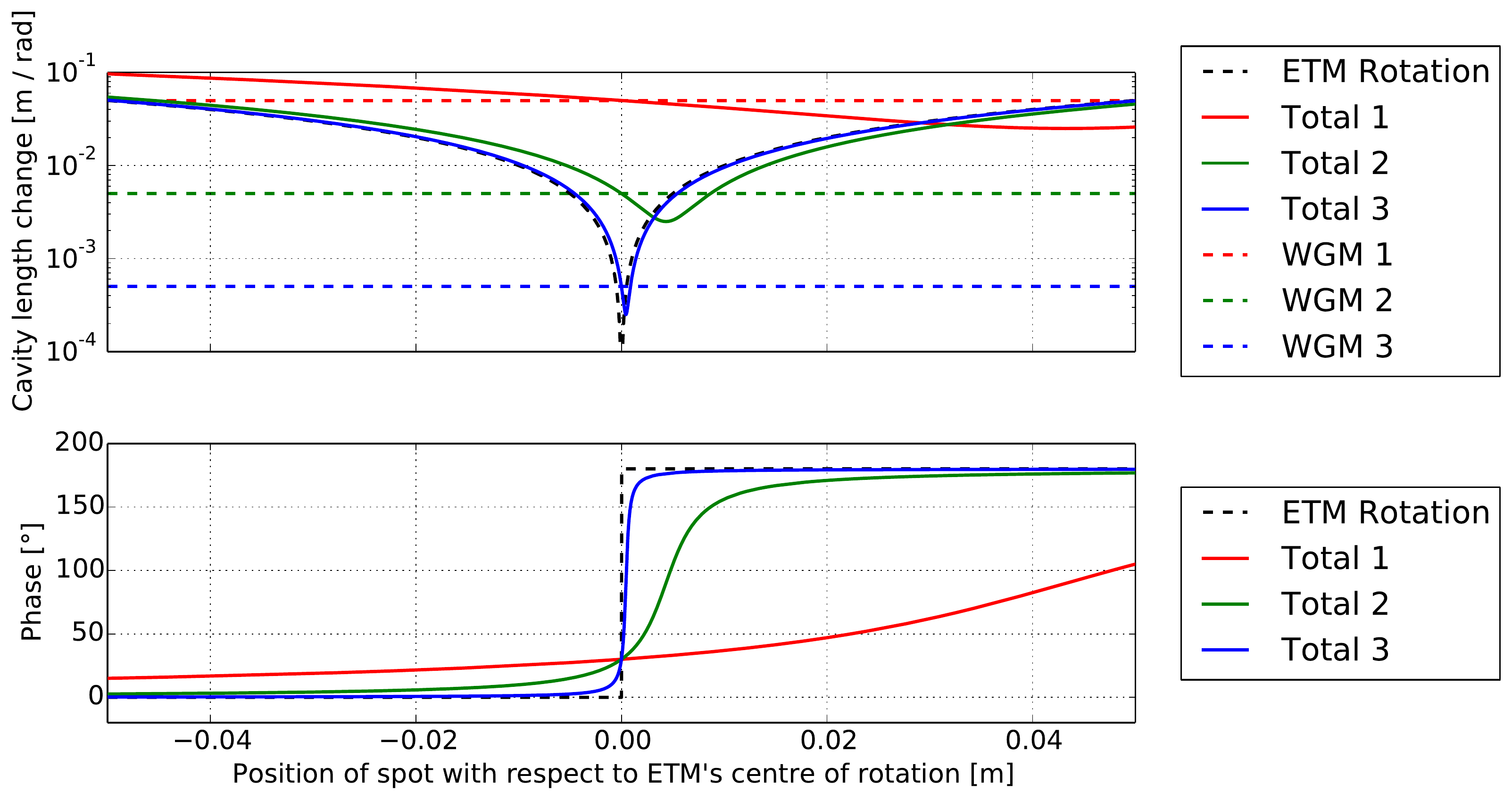}
    \caption{Simulations of indicative cavity longitudinal error signals during ETM rotation for different levels of WGM coupling. The signals are functions of the transverse position of the reflected light relative to the ETM's centre of rotation, the angle of rotation, the mirror depth and the WGM's coupling level. The rotation to longitudinal coupling of the ETM (black dashed line) combines with the transverse to longitudinal coupling of the WGM (red, green and blue dashed lines) to produce cavity length changes (red, green and blue solid lines). In this example configuration, the ETM rotation is \SI{1e-7}{\radian}, the ETM's depth is \SI{0.1}{\meter} and the corresponding WGM coupling levels are 1:370 (red), 1:3700 (green) and 1:37000 (blue).}
    \label{fig:individual_factors}
  \end{center}
\end{figure}

Considering the ETM's level of rotation and its dimensions and mass, it is possible to calculate the cavity length change due to the two geometrical effects shown in Equation~\ref{eq:etm_length_change} and then, from the residual cavity length change, infer the WGM's coupling level. The phase effect associated with transverse to longitudinal coupling is expected to be independent of spot position, whereas there is a phase change about the ETM's centre of rotation. It is therefore expected that a spot position will exist, for a non-zero WGM transverse coupling level, offset from the ETM's centre of rotation, for which there is a cavity error signal minimum. This effect arises as a result of $\delta l_W$ and $\delta l_{E}$ combining coherently (see
Figure~\ref{fig:individual_factors}). The spot position corresponding to the cavity error signal minimum allows the WGM's transverse to longitudinal coupling level to be inferred.

Examples of WGM coupling levels yielding cavity length changes smaller than (blue), larger than (red) and roughly equivalent to (green) the ETM's effects are shown in Figure
\ref{fig:individual_factors}. For cases where the WGM's coupling level yields a significant cavity length change with respect to that of the ETM's rotation, coherent combination creates a trough offset from the ETM's centre of rotation.

\subsection{The Glasgow 10~m Prototype}
\label{sec:glasgow10m}

The Glasgow\,\SI{10}{\meter} prototype facility provided a test bed in which the WGM's
transverse to longitudinal coupling could be quantified. The prototype is housed
in a Class 1000 clean room and consists of an input bench at atmospheric
pressure and a vacuum envelope able to reach pressures of order $10^{-5}$\,mBar. The envelope consists of nine \SI{1}{\meter} diameter steel tanks,
each connected by steel tubes, arranged into two parallel arms of length \SI{10}{\meter}, with a shorter arm for input optics situated between them.

\begin{figure}
  \begin{center}
   
\includegraphics[width=0.9\columnwidth]{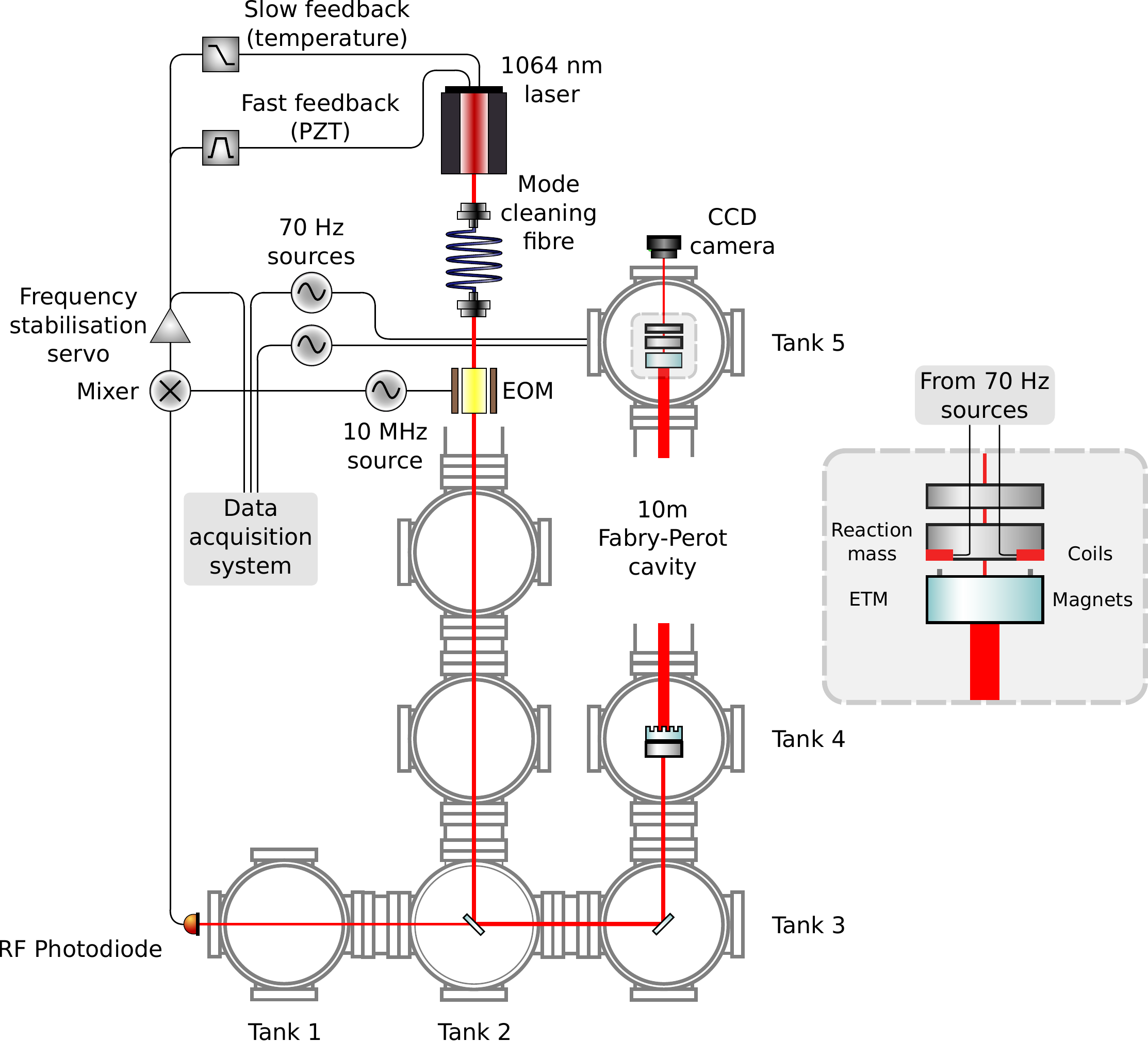}
    \caption{The experimental setup in the prototype facility.
The laser light is passed through input optics (not shown), a mode cleaning fibre and an EOM before being coupled into the vacuum system \emph{via} a periscope. It then travels to tank 2 where it is reflected off a beam splitter and directed into one of the arms of the prototype by a steering mirror in tank 3. The two cavity mirrors in tanks 4 and 5 form a \fp cavity. The cavity mirrors are suspended from triple stage suspensions, and the beam splitter and steering mirror are both suspended from double suspensions. \\
\\The ETM is rotated in yaw using the \SI{70}{\hertz} source. It is fed to a coil driver where it is coupled into tank 5 \emph{via} a vacuum feedthrough. Coil formers on the front edges of the reaction mass contain wound copper wire connected to the vacuum feedthrough. Magnets are attached to the back of the ETM. The reaction mass is behind the ETM, containing a hole in its centre to allow light to exit the vacuum tank where it can be viewed with the CCD camera. A larger version of the contents of tank 5 can be viewed in the panel to the right of the figure. \\
\\The cavity is held on resonance by the frequency stabilisation servo. This feeds back to the light's frequency \emph{via} the laser crystal's temperature below \SI{12}{\hertz} and its PZT above \SI{12}{\hertz} up to a unity gain frequency of \SI{14}{\kilo \hertz}.}
    \label{fig:prototype_setup}
  \end{center}
\end{figure}

In the experiment, \SI{1064}{\nano\meter} laser light was passed through a single-mode fibre to provide spatial filtering and an electro-optic modulator (EOM) to impose RF sidebands on the light to facilitate PDH control. The light was then coupled into the vacuum system \emph{via} a periscope. This configuration can be viewed in Figure~\ref{fig:prototype_setup}.

Tanks 2 and 3 housed a beam splitter and steering mirror, respectively, attached to double stage suspensions. In tanks 4 and 5 were sets of two triple suspension chains based on the GEO-600 design \cite{Plissi2000}. A viewport present to the rear of tank 5, and to the side of tank 1, allowed for light to exit the vacuum envelope for the purposes of sensing and control.

\begin{figure}
  \begin{center}
   
\includegraphics[width=0.3\columnwidth]{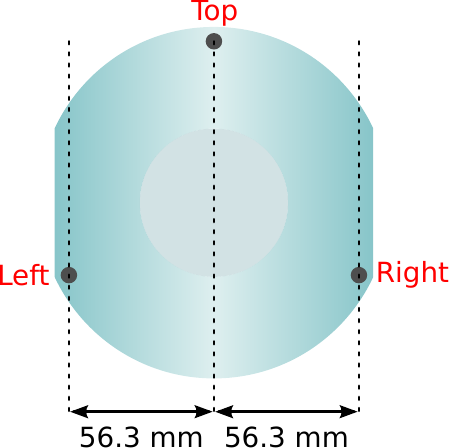}
    \caption{The positions of the magnets on the rear surface of the ETM. The magnet designations used in this article are shown in red text. The top magnet is positioned at the centre of yaw, near the top of the mass. The left and right magnets are positioned \SI{56.3}{\milli \meter} either side of the centre of yaw. Coils on the ETM's reaction mass (not shown) are positioned coaxially behind each magnet.}
    \label{fig:etm_rear}
  \end{center}
\end{figure}

The WGM was attached to an aluminium block of mass \SI{2.7}{\kilo\gram} and suspended from tank 4's cascaded (triple) pendulum, forming the cavity's ITM. A silica test mass, also \SI{2.7}{\kilo\gram}, with a \SI{40}{ppm} transmission coating, was used as the ETM, suspended from a similar triple pendulum in tank 5. On the rear surface of the ETM were three magnets for the purpose of actuation, the positions of which are shown in Figure~\ref{fig:etm_rear}. With optimal alignment the mirrors formed an overcoupled cavity with finesse \num{155}.

A three-stage reaction chain was placed behind the triple pendulum of the ETM to provide voice coil actuation upon the magnets on the ETM's rear surface. The upper and intermediate stages were identical to those of the chain carrying the ETM, however\textemdash for the purposes of another experiment, not reported here\textemdash the lower stage was split into two parts separately suspended from the intermediate stage. The part closer to the ETM was a \SI{1.8}{\kilo\gram} aluminium block that carried the voice coils. The other part was a \SI{0.9}{\kilo\gram} aluminium block required to balance the suspension.

\begin{table}
  \begin{center}
     \begin{tabular}{|l|l|}
       \hline
       \textbf{Parameter}        & \textbf{Description}	\\ \hline
         Cavity input power      & Approx. \SI{150}{\milli\watt} \\ \hline
         ETM transmissivity      & $40$~ppm \\ \hline
         ETM radius of curvature & \SI{15}{\meter} \\ \hline
         ETM spot size           & \SI{2.138}{\milli \meter} \\ \hline
         ITM transmissivity      & \SI{4}{\%} \\ \hline
         ITM radius of curvature & $\infty$ \\ \hline
         ITM spot size           & \SI{1.554}{\milli \meter} \\ \hline
         Cavity length           & \SI{9.81}{\meter} \\ \hline
         Cavity finesse          & \SI{155}{} \\ \hline
         Cavity g-factor         & \SI{0.347}{} \\ \hline
         Beam waist size         & \SI{1.554}{\milli \meter} \\ \hline
         Beam waist position     & At ITM \\ \hline
         Sideband frequency      & \SI{10}{\mega\hertz} \\ \hline
     \end{tabular}
     \caption{Cavity parameters.}
     \label{tab:cavity_parameters}
  \end{center}
\end{table}

\subsection{Measuring Cavity Length Changes}

An RF photodetector was placed at the viewport on tank 1, where it could view
the light reflected from the cavity. By using PDH demodulation, the signal from
this photodetector provided an error signal for the cavity length. This signal
was fed back to the laser \emph{via} the frequency stabilisation servo to maintain cavity resonance. The frequency stabilisation servo's high frequency feedback signal\textemdash a voltage applied across the laser's piezoelectric transducer (PZT)\textemdash provided a means of calibrating cavity length changes at frequencies greater than \SI{12}{\hertz}. Using the PZT's frequency response, \SI{1.35}{\mega\hertz \per \volt_{rms}}, the cavity length change $\delta l$ per error signal volt could be calculated to be \SI{133}{\nano\meter \per \volt_{peak}}.

\section{Measurements and Analysis}
\label{sec:measurements}

From the orientation of the WGM's gratings, it was expected that actuation of the ETM in yaw, which would scan the cavity light across the WGM's surface transverse to the direction of its grooves, would exhibit WGM transverse to longitudinal coupling if present.

For the purposes of actuation upon the ETM, two sinusoidal signals $V_L$ and $V_R$ (corresponding to the left and right voice coils on the ETM's reaction mass, respectively) were produced using separate, phase locked signal generators. A signal frequency of \SI{70}{\hertz} was chosen so as to be above the suspensions' pole frequencies but low enough to provide an adequate signal-to-noise ratio. The signals $V_L$ and $V_R$, with suitable balancing (see below), could then be actuated in- or out-of-phase to produce longitudinal or yaw actuation upon the ETM, respectively.

When $V_L$ and $V_R$ were identical in magnitude but out-of-phase, the ETM's movement contained a linear combination of rotational and longitudinal components due to force imbalances between the voice coils. To ensure that actuation upon the ETM contained only a yaw component, the cavity's longitudinal error signal was minimised during out-of-phase actuation by changing the gain of $V_L$. This balanced the magnitude of the torque applied by each actuator to the left and right sides of the ETM. Any WGM transverse to longitudinal coupling present would act with phase orthogonal to this voice coil actuation and would thus be unchanged by the torque balancing.

Pitch actuation upon the ETM, which would scan the cavity light in a direction parallel to the WGM's grooves, was not expected to contribute to the cavity's error signal via the WGM's coupling. However, unintended pitch actuation upon the ETM would couple into the cavity's length \emph{via} the same geometrical mechanism as yaw shown in Equation~\ref{eq:etm_length_change}. To minimise the ETM's pitch component during actuation in yaw, the cavity's error signal was minimised by applying an offset voltage to the top coil. In practice, minimal pitch coupling was achieved when the offset signal was zero.

\subsection{Actuator Calibration}

To calibrate the cavity's longitudinal response to voice coil actuation, the voice coils were actuated with the balanced $V_L$ and $V_R$ signals in-phase at a frequency $f = \SI{70}{\hertz}$ for a period of \SI{120}{\second}. This, along with the ETM's mass $m$, could then be used to obtain the force applied to the ETM by the voice coils:
\begin{equation}
  F = 4 \pi^2 f^2 m \delta l.
  \label{eq:force_calibration}
\end{equation}

\subsection{Measurement of Waveguide Mirror Transverse to Longitudinal Coupling}
\label{sec:length_changes}

Four spot positions corresponding to $y$ in Equation~\ref{eq:offset_effect} were chosen across the surface of the ETM. The input beam was aligned to the cavity axis corresponding to each spot position using the beam splitter and steering mirror nearest to the ITM, and the cavity mirrors were aligned to create a fundamental mode resonance. The voice coil signals $V_L$ and $V_R$ were set out-of-phase to produce motion on the ETM in yaw. The magnitudes of $V_L$ and $V_R$ were not altered between the longitudinal calibration and this yaw actuation, so it was expected that the previously outlined minimisation of yaw to tilt actuation would also result in minimal longitudinal to tilt actuation. The cavity length signal was recorded for a period of \SI{300}{\second}.

For each nominal spot position an additional measurement was taken with $V_L$ set to $\pm \SI{0.1}{\volt}$ from its balanced setting for a period of \SI{60}{\second}. This allowed two additional data points to be obtained for each spot position. By calculating the gradient (cavity length change per spot position with respect to the centre of yaw) of the central and inner-left spot positions, it was possible to assign an effective spot position for each of the offset points.

The spot positions used to obtain cavity error signals are shown in Table~\ref{tab:spot_positions}. These positions are shown with respect to the centre of the ETM's reflective surface. The spot positions were subject to two sources of error: the measurement of the spot positions with respect to the centre, and the error in the ETM's centre of rotation due to misalignment between the voice coils and their corresponding magnets. The spot position error was assumed to be +/-\SI{1}{\milli\meter} from visual inspection of the suspensions, measured \emph{via} the CCD camera placed in transmission of the ETM, using the known width of the ETM's reaction mass as a calibration. The error in the spot position measurements dominated the error in voice coil alignment. Although misaligned voice coils could have lead to a change in the expected ETM force coupling (leading to a change in the centre of rotation of the ETM), it was found from separate measurements that the effect of any possible misalignment during the experiment could only account for a drop in force of \SI{0.11}\%. This contributed a negligible error (+/-\SI{0.03}{\milli\meter}) to the results.

\begin{table}
  \begin{center}
     \begin{tabular}{|c|c|c|}
       \hline
       \multicolumn{3}{|c|}{\textbf{Spot position [\SI{}{\milli \meter}]}} \\ \hline
       \SI{-0.1}{\volt} & \SI{0.0}{\volt} & \SI{+0.1}{\volt} \\ \hline\hline
       \num{-12.9} & \num{-12.5} & \num{-12.1} \\ \hline
       \num{-5.4} & \num{-5.0} & \num{-4.6} \\ \hline
       \num{-0.4} & \num{0.0} & \num{0.4} \\ \hline
       \num{12.1} & \num{12.5} & \num{12.9} \\ \hline
     \end{tabular}
     \caption{Spot positions on the ETM for the far left, inner left, central and right positions, respectively. The positions are shown in groups of three corresponding to the offset applied to $V_L$. All spot positions have an error of +/-\SI{1}{\milli\meter}.}
     \label{tab:spot_positions}
  \end{center}
\end{table}

Knowledge of the distance of the ETM's voice coils from the centre of rotation, $y_c$; the ETM's moment of inertia, $I$; the coil driving frequency, $f$; and the force
calibration from Equation~\ref{eq:force_calibration}, allowed the rotation angle
to be obtained geometrically using the relation
\begin{equation}
  \theta = \frac{F y_c}{4 \pi^2 f^2 I}.
  \label{eq:rotation_calibration}
\end{equation}
The numerical simulation tool \emph{FINESSE} \cite{Freise2004} was then used to calculate $\kappa$ for the cavity parameters shown in Table~\ref{tab:cavity_parameters}. This was determined to be \SI{18.5}{\meter \per \radian}. The WGM's transverse displacement was then the product of $\kappa$ and $\theta$.

\subsection{Analysis of the Coupling Level}
\label{sec:simulations}
Using the known contribution to the cavity length signal from the rotation of the ETM, $\delta l_E$, and the cavity length signals $\delta l$ measured during the experiment, the WGM's coupling level could be calculated statistically using Bayes' theorem. For this experiment, Bayes' theorem can be expressed mathematically as:
\begin{equation}
  p \left( \vec{\omega} | \mathcal{D} \right) \propto p \left( \mathcal{D} | \vec{\omega} \right) p \left( \vec{\omega} \right),
  \label{eq:bayes}
\end{equation}
where $p \left( \vec{\omega} | \mathcal{D} \right)$ is the probability density distribution of the experimental parameters, $\vec{\omega}$, given the observed data, $\mathcal{D}$ (the \emph{posterior}); $p \left( \mathcal{D} | \vec{\omega} \right)$ is the likelihood and $p \left( \vec{\omega} \right)$ is the probability distribution of the experimental parameters. The observed data $\mathcal{D}$ are the measured cavity error signals for each of the spot positions.

In this analysis we are primarily interested in estimates of the model parameters. We are therefore free to ignore the constant evidence factor $p \left( \mathcal{D} \right)$ present in Bayes' theorem when calculating the posterior. In the future it may be of interest to compare different models for the coupling level (or lack thereof), in which case the evidence could be calculated to obtain a model odds ratio.

\subsubsection{Model and Parameters}
To obtain a posterior for the WGM's coupling level, it was necessary to build a model and state prior belief of the parameters' probability distributions.

In the model, the ETM's geometrical longitudinal effect at arbitrary spot position $y$ (Equation~\ref{eq:etm_length_change}) for the rotation and mirror depth used in the experiment was combined coherently with a specified level of WGM transverse to longitudinal coupling, $\omega_1$. It was then possible to predict the total change in cavity length $\delta l$ as a function of spot position $y$, given the fixed parameters $\theta$, $\kappa$ and $d$, using equations~\ref{eq:wgm_length_change} and \ref{eq:etm_length_change}:
\begin{equation}
  \begin{split}
    \delta l \left( \vec{\omega}, y, \theta, \kappa, d \right) & = \delta l_W \left( \theta, \kappa, \omega_1 \right) + \delta l_E \left( y, \theta, d \right) \\
    & \approx \theta \kappa \omega_1 + y \theta + \frac{d}{4} \theta^2.
  \end{split}
\end{equation}

The effect of \emph{beam smearing} was also considered. The suspended optics contain residual displacement noise, leading to a broadening of the trough at which the ETM's longitudinal coupling and any WGM coupling cancel (see Figure~\ref{fig:individual_factors}). To model this effect, the assumption was made that the motion of the spots on the ETM followed a Gaussian distribution about their nominally measured position. Eight-hundred small `offset distances' $\delta y$ were applied uniformly to the spot positions, drawn from a randomly generated Gaussian distribution. The number of offset distances was chosen as a compromise between adequate statistical significance and technical constraints. Calculating the cavity length change as a function of spot position for each of these offset positions, and combining them in an uncorrelated sum, allowed an average, `smeared' signal to be modelled which more closely resembled the measurements. The standard deviation of the Gaussian distribution was an additional parameter, $\omega_2$, provided as an input to the model.

The summing of signals introduced by the modelling of beam smearing led to an artificial increase in the magnitude of the model's predicted cavity length signals. To compensate for this effect, a further parameter was introduced: a multiplicative scaling factor, $\omega_3$, applied uniformly to the model. This factor also had the additional effect of compensating for the uncertainty in the calibrated cavity length signals. By marginalising over a suitable distribution of scaling factors, it was possible to account for this uncertainty in the analysis of the WGM's coupling level. The model used in the analysis to predict the smeared, scaled cavity length change, $\delta l'$, was then:
\begin{equation}
  \delta l' \left( \vec{\omega}, y, \theta, \kappa, d \right) = \omega_3 \sqrt{\sum_{i=1}^{800} \delta l \left( \vec{\omega}, y + \delta y_i, \theta, \kappa, d \right)^2},
  \label{eq:model}
\end{equation}
where $\delta y_i$ is the $i^\text{th}$ offset distance, drawn from a Gaussian distribution with standard deviation $\omega_2$.

\subsubsection{Likelihood}
The likelihood function assumed for the model was a Gaussian distribution,
\begin{equation}
  p \left( \vec{\omega} | \mathcal{D} \right) \propto \exp \left( -\frac{1}{2} \sum_{i=1}^{N} \frac{\left( \mathcal{D}_i - \delta l' \left( \vec{\omega}, y_i, \theta, \kappa, d \right) \right)^2}{\sigma^2} \right),
  \label{eq:likelihood}
\end{equation}
where $N$ is the number of spot positions and $\sigma^2$ is the (identical) variance of each of the measured spot positions.

\subsubsection{Priors}
Bayes' theorem requires an assumption of probability distributions (\emph{priors}) for each of the free parameters prior to the consideration of the measured data. The assumptions made for each free parameter in the model can be found in Table~\ref{tab:priors}. The upper bound on coupling was assumed to be a factor \num{10} better than the grating mirror measured in~\cite{Barr2011}, given the indication from~\cite{Brown2013} that no coupling is present. The bounds on the scaling factor and spot smearing standard deviation were chosen from earlier observations of the behaviour of the signals during the experiment. All priors were assumed to be uniform.

\begin{table}
  \begin{center}
     \begin{tabular}{|p{3cm}|c|c|c|}
       \hline
       \textbf{Parameter}   & \textbf{Symbol}     & \textbf{Distribution} & \textbf{Dimensions} \\ \hline
       WGM transverse to longitudinal coupling & $\omega_1$ & Uniform, $\left[ 0, \frac{1}{1000} \right]$ & $\frac{\SI{}{\meter} \text{ (longitudinal)}}{\SI{}{\meter} \text{ (transverse)}}$ \\ \hline
       Spot smearing noise standard deviation & $\omega_2$ & Uniform, $\left[ 0, 3 \times 10^{-3} \right]$ & $\SI{}{\meter} \text{ (transverse)}$ \\ \hline
       Calibration scaling                    & $\omega_3$ & Uniform, $\left[ 0, \frac{1}{10} \right]$ &  \\ \hline
     \end{tabular}
     \caption{The distributions assumed for each of the free parameters in the model, along with their dimensions, prior to the computation of the posterior.}
     \label{tab:priors}
  \end{center}
\end{table}

\subsubsection{Algorithm}
A form\footnote{\emph{``Yet Another Matlab MCMC code''} by Matthew Pitkin. Available as of time of writing at \url{https://github.com/mattpitkin/yamm}.} of the Metropolis-Hastings Markov-Chain Monte-Carlo (MCMC) algorithm \cite{Hastings1970} was applied to the model to marginalise over the three parameters. The outputs of the MCMC are a chain of samples (values at each parameter) that are drawn from the posterior distribution. A histogram of samples for a given parameter gives the marginal posterior distribution for that parameter from which the mean and standard deviation can be calculated.

To ensure the convergence of the MCMC on the posterior, a `burn-in' period of \num{100000} iterations was performed. The convergence was verified manually following completion. A further \num{100000} iterations were then used to sample from the posterior and this second set is the one that we used for our results.

\section{Results}
\label{sec:summary}
\begin{figure}[H]
  \begin{center}
   
\includegraphics[width=\columnwidth]{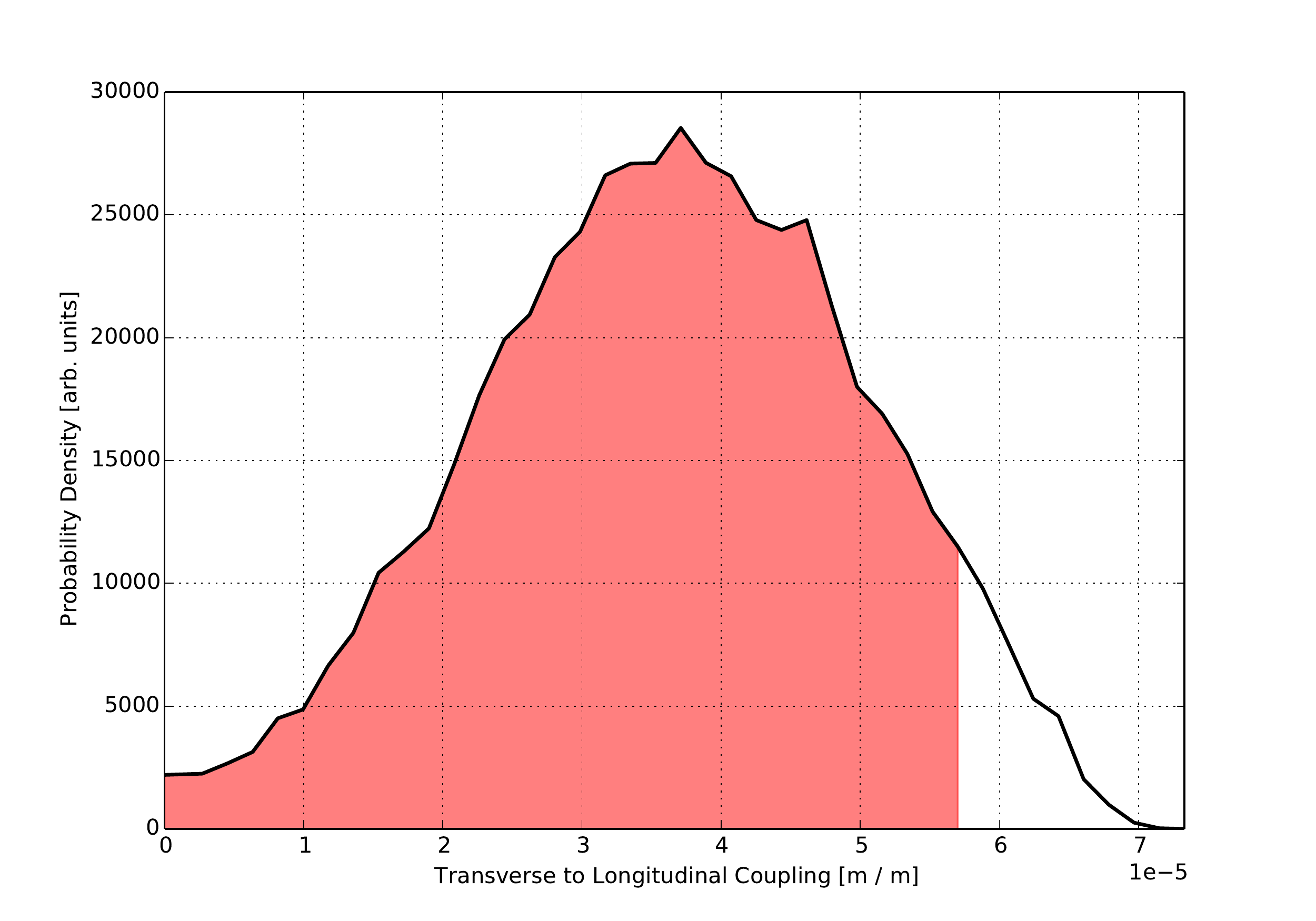}
    \caption{Posterior probability density distribution of WGM coupling levels (in units of meters longitudinal per metre transverse) yielded by statistical analysis of the data. The red shaded region shows the coupling levels falling within the most probable 95\% of the distribution.}
    \label{fig:final_result_prob}
  \end{center}
\end{figure}

\begin{figure}[H]
  \begin{center}
   
\includegraphics[width=\columnwidth]{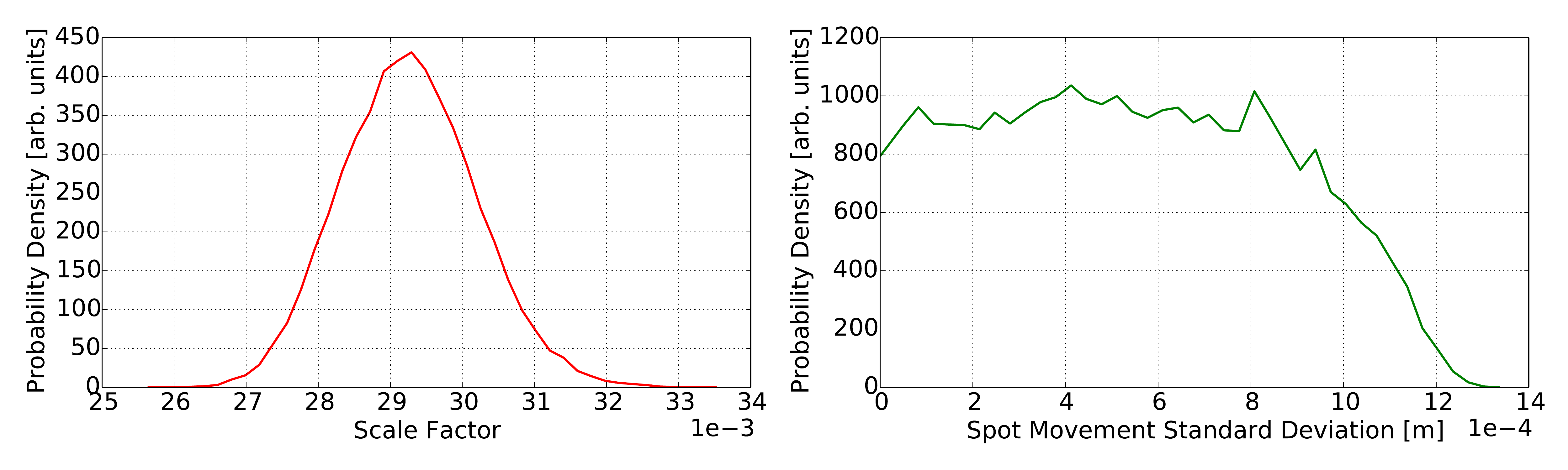}
    \caption{Posterior probability density distribution of other parameters used in the analysis: scaling applied to the model's predicted longitudinal signal (left plot) and the standard deviation assumed for the Gaussian distribution used to model beam smearing (right plot). Both distributions lie well within their prior ranges (see Table~\ref{tab:priors}).}
    \label{fig:final_posteriors}
  \end{center}
\end{figure}

From the parameter marginalisation it was possible to produce a posterior probability density distribution for the coupling level as shown in Figure~\ref{fig:final_result_prob}. The coupling level predicted from the distribution is bounded between 0 and 1:17000 with 95\% confidence, with a mean coupling level of 1:27600. The probability density distributions for the scaling and standard deviation parameters are shown in Figure~\ref{fig:final_posteriors}. The scaling posterior distribution indicates a mean value of \num{29.3e-3} with standard deviation \num{0.94e-3}. The posterior distribution for the beam smearing parameter indicates a range of possible values between \num{0} and \SI{1.3e-3}{\meter}.

\begin{figure}[H]
  \begin{center}
   
\includegraphics[width=\columnwidth]{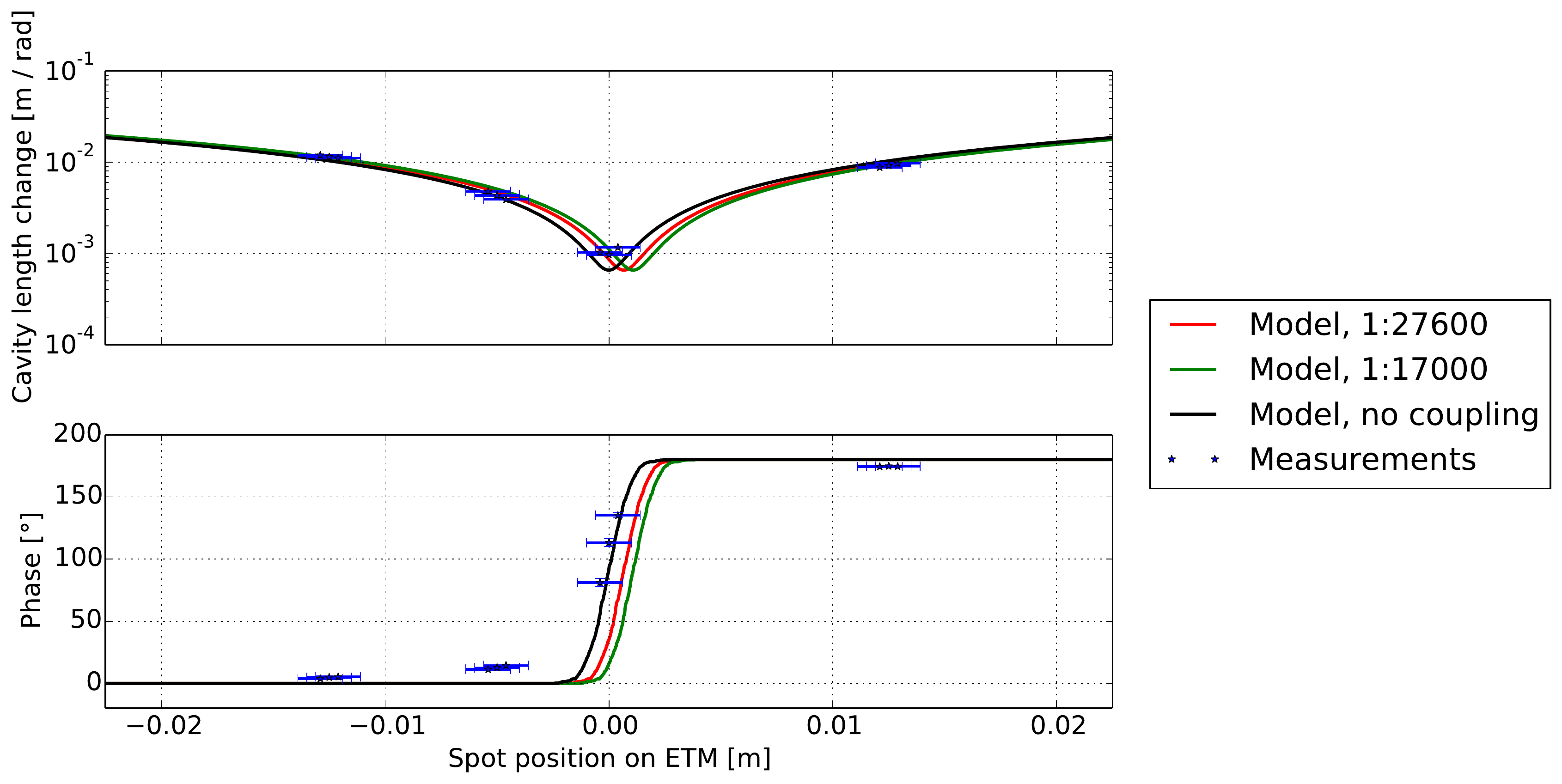}
    \caption{Measurements and simulations of the cavity length signal for spot positions with respect to the ETM's centre of yaw. The calibrated cavity length change per radian (vertical axis) from the measurements is shown (blue stars) alongside the model's simulated cavity length changes per radian for the mean (red), 95\% upper limit (green) and zero (black) WGM coupling levels. The simulated plots use a scaling factor of \num{29.3e-3} and a beam smearing standard deviation of \SI{0.8e-3}{\meter}.
    \bigskip
    \\ Error bars are shown on the measured spot positions corresponding to their uncertainty. The errors in cavity length change are obtained from the noise floor surrounding each measurement. The noise floors were approximately constant for all measurements, with mean value \SI{8e-5}{\meter \per \radian}. Phase error bars are visible for the central values. The errors on each phase measurement, from left to right, are: +/-\num{0.0188}, +/-\num{0.0254}, +/-\num{0.0283}, +/-\num{0.1387}, +/-\num{0.1721}, +/-\num{0.2178}, +/-\num{3.2726}, +/-\num{3.2303}, +/-\num{2.0603}, +/-\num{0.0385}, +/-\num{0.0342} and +/-\num{0.0336} degrees.}
    \label{fig:final_result}
  \end{center}
\end{figure}

The measured cavity length signals as well as the 95\% upper limit and mean WGM coupling level predicted by the analysis are shown in Figure~\ref{fig:final_result}. The phase discrepancy between the model and the measurements, as witnessed in this figure most profoundly for the spot positions around \SI{-5e-3}{\meter}, is thought to be an artefact from the modelling of the beam smearing effect. The residual test mass motion that motivated the inclusion in the model of beam smearing may have contained some non-Gaussian behaviour.

The upper limit on the predicted coupling level, 1:17000, represents a significant improvement over previously measured grating designs such as the \nth{2} order Littrow grating measured in \cite{Barr2011}, where the coupling factor was of order 1:100.

\subsection{Acknowledgements}
The authors would like to thank members of the LIGO Scientific Collaboration for fruitful discussions. The Glasgow authors are grateful for the support from the Science and Technologies Facility Council (STFC) under grant number ST/L000946/1. The Jena authors are grateful for the support from the Deutsche Forschungsgemeinschaft under project Sonderforschungsbereich Transregio 7.

\printbibliography

\end{document}